# Periodic Variables and Gyrochronology in the Open Cluster NGC 2301

TUGULDUR SUKHBOLD

Steward Observatory, University of Arizona, Tucson, AZ 85721; tuguldur@email.arizona.edu

STEVE B. HOWELL

National Optical Astronomy Observatory, Tucson, AZ 85719; howell@noao.edu

**ABSTRACT.** We present the results of a search for periodic variables within 4078 time-series light curves and an analysis of the period-color plane for stars in the field of the open cluster NGC 2301. One hundred thirty-eight periodic variables were discovered, of which five are eclipsing binary candidates with unequal minima. The remaining 133 periodic variables appear to consist mainly of late-type stars whose variation is due to rotation modulated by star spot activity. The determined periods range from less than a day to over 14 days and have nearly unreddened $B - R$ colors in the range of 0.8 to 2.8. The Barnes (2003) interpretation of the period-color plane of late type stars is tested with our data. Our data did not show distinct I and C sequences, likely due to nonmember field stars contaminating in the background, as we estimate the total contamination to be 43%. Using different assumptions, the gyrochronological age of the cluster is calculated to be $210 \pm 25$ Myr, which falls in the range of age values (164–250 Myr) determined by previous studies. Finally, we present evidence which nullifies the earlier suggestion that two of the variable stars in NGC 2301 might be white dwarfs.

*Online material:* color figures

## 1. INTRODUCTION

The young metal-rich (Fe/H = 0.06) open cluster NGC 2301 ($\alpha = 0.6^h 51^m 48^s.00$, $\delta = +00^h 28^m 00^s.0$, J2000) is located 872 pc away toward the constellation of Monoceros (metallicity and location according to the open cluster database WEBDA),[1] almost lying on both galactic and celestial equators, but with an amazingly low reddening value of $E(B - V) = 0.028$ mag. In the last 50 years, only a few papers have appeared that discuss the proper motion-based member probabilities and some static photometry of stars in the cluster. A pioneering study on membership probabilities, the earliest paper by Plaut et al. (1959) determined proper motions for 98 stars in the region using three photographic pairs of plates with epoch differences of 37.80 yr, 58.08 yr, and 54.11 yr. Later, Aiad (1986) obtained proper motions of 190 stars again with photographic plates, having an epoch difference of 75.72 yr. Early photographic photometry in the cluster was conducted by Grubissich & Purgathofer (1962) and subsequent investigations based on photoelectric photometry are discussed in papers of Mohan (1988) and Kim et al. (2001). In 2004 February, a 14 night-long observing run was used to collect high-precision CCD time-series photometry of the open cluster NGC 2301 (Howell et al. 2005, hereafter H05). These observations made use of the OPTIC imager (see Howell et al. 2003) on the University of Hawaii 88″ telescope. The best precision of the CCD photometry was ~1–2 mmag per image; it covered a magnitude range of $+10 \leq R \leq +19.5$, and measured point sources of color range $-0.5 \leq B - R \leq 3.7$. The photometric depth and precision presented in H05 and in this article are beyond any former work on NGC 2301, and only a very few of our observed cluster stars overlap with previous studies.

In this article, we present the results of an analysis of the 140 variable stars, from the 4078 discussed in H05 and Tonry et al. (2005), which are shown to be, at a 99% or greater probability level, periodic variables. More information on the observational techniques used and the initial variability statistics for these data are discussed in Tonry et al. (2005). As to membership, there is no information available from previous studies for stars fainter than $R = 12$. Thus, we attempt to suggest membership based on spatial location of the variables in both the cluster color-magnitude diagram (CMD) and celestial coordinates relative to the accepted cluster radius and the cluster center.

The discovery of eclipsing binaries (EBs) in open clusters is of interest since these stars are excellent sources to provide independent distance and other information for the cluster as well as fundamental determinations of the binary components' radii, mass, and luminosities (see, e.g., Guinan et al. 1998, Ribas et al. 2005). EBs have also been used to determine accurate distances to local group galaxies such as the LMC (e.g., Fitzpatrick et al.

---

[1] At http://www.univie.ac.at/webda/.

2003; Ribas 2004), SMC (e.g., Hilditch et al. 2005), M31 (e.g., Ribas et al., 2005), and M33 (e.g., Bonanos et al., 2006). In order to make such determinations and use EBs as "standard candles," good spectroscopic observations must be conducted in order to determine the radial velocity motions for the component stars.

The results of the light-curve analysis herein show that most of the periodic variables are rotating stars. Analyzing rotation periods of late-type stars in young clusters, Barnes (2003, hereafter B03) presented his own interpretation of the rotational evolution of cluster members. The reader should also see Bouvier et al. (1997) and Irwin et al. (2007) for alternate approaches; for a general review, see Irwin & Bouvier (2009). B03 proposed two rotational sequences of stars based on separate groups of stars differing from each other by their dependencies of rotation on mass, and an interpretation framework that "explains" the evolution of these sequences. B03 also analyzed how the locations of these sequences depend on the age of the cluster and proposed an empirical model which could be used for the age determination of the cluster. Meibom et al. (2008, hereafter M08) obtained high-quality photometric and spectroscopic data on the 150 Myr open cluster M35, and analyzed and applied the B03 interpretation in great detail. In this article, we present the analysis of the period-color plane for our newly found periodic variables in the field of NGC 2301.

The candidate rotating and EBs presented here start the process toward a deeper understanding of NGC 2301. Spectroscopic observations can now further the process using the information we provide.

## 2. LIGHT-CURVE ANALYSIS

H05 examined the light curves of 4078 stars in the field of NG C2301, and at a conservative level, it was estimated that ~56% of these (~2300) are variable. H05 used a reduced chi-squared value for a constant fit to the full light curve, assumed that a value of $X^2/n = 1.0$ would represent a constant star, and took a value of $X^2/n = 3.0$ or greater to indicate variability at a conservative level. Period searching from 30 minutes to 14 days was performed using the PDM technique (Stellingwerf 1978) in Tonry et al. (2005). Herein each light curve with a 99% or greater probability of being periodic was phased and examined by eye. This amounted to visual inspection of 140 time-series phased light curves. We note that many of the periodic variables we discovered have shallow variations, and for the EB candidates, nearly equal primary and secondary eclipse depths. Due to this effect, the period-finding routine often found a "best-fit" period that was really one-half of the true period as noted by visual inspection. We also looked for deeply eclipsing binaries indicated by a light-curve drop of greater than 1 mag. None of these were found. A total of 138 highly confident periodic variables were culled from the original candidate list. The newly discovered periodic variables are listed in Tables 1 and 2. Stars fainter than ~17.5 in our sample provide poorer photometry with correspondingly poorer statistics for robust periodicity determination. Additionally, stars fainter than 17.5 are more difficult to follow up spectroscopically. Thus, we have used a magnitude-limited sample ($R$ brighter than 17.5, for the most part) for our examination of light curves of periodic variables and the gyrochronological age determination of NGC 2301.

The five stars in Table 1 have light curves that provide evidence for candidacy as eclipsing binaries; two unequal minima, a usual signature of an eclipsing binary system. We plot their phased light curves in Figure 1. The orbital periods found are short, all less than 1 day. In any time-limited observational program, shorter period sources will be preferentially discovered.

Our four EB candidates will be the most likely systems to be used in distance determination studies. Of course there are other types of eclipsing systems that do not necessarily show two minima or two unequal minima. Few systems such as these (e.g., W UMa-type) are likely to exist in our sample of additional periodic variables. We note that the photometric study conducted on the central region of NGC 2301 by Kim et al. (2001) found 9 variables including 5 EBs. The paper by Kim et al. (2001) did not provide R.A. and declination values for their new variable stars, but only an image with the variable stars marked on it. Thus, when H05 attempted to cross reference the 9 Kim et al. (2001) stars to their variables, they were able to uniquely match only 7 of them. One of the matched stars is equal to our discovered eclipsing binary C163, and our orbital period ($P = 0.250$ days) agrees with the listed value of Kim et al. (see Table 1).

Table 2 lists the additional 133 periodic variables we have found, of which four example light curves (S210, W163,

TABLE 1
FIVE ECLIPSING BINARY CANDIDATES

| ID | R.A. | Decl. | Period (days) | Amplitude (mag) | $R$ (mag) | $B - R$ (mag) | RV (km s$^{-1}$) |
|---|---|---|---|---|---|---|---|
| C163 ............ | 102.9303 | 0.52185 | 0.25 | 7.89E-02 | 15.404 | 1.777 | 191 |
| N157 ............ | 103.1275 | 0.60832 | 0.256 | 6.63E-01 | 16.322 | 1.771 | 189 |
| SE271 ............ | 103.1925 | 0.36508 | 0.334 | 5.19E-01 | 17.242 | 2.154 | 168 |
| SE292 ............ | 103.1825 | 0.28021 | 0.281 | 1.91E-01 | 17.386 | 1.834 | 182 |
| W235 ............ | 102.8222 | 0.582362 | 0.31 | 1.05E-01 | 16.62 | 1.908 | 176 |

TABLE 2
THE 133 ADDITIONAL PERIODIC VARIABLES

| ID | R.A. | Decl. | Period (days) | Amplitude (mag) | $R$ (mag) | $B - R$ (mag) |
|---|---|---|---|---|---|---|
| C79 | 102.9266 | 0.425409 | 2.670 | 3.07E-02 | 13.780 | 0.904 |
| C85 | 102.993 | 0.44713 | 2.519 | 5.17E-02 | 13.846 | 0.934 |
| C96 | 103.0153 | 0.501051 | 4.834 | 4.40E-02 | 13.688 | 0.949 |
| C112 | 103.0432 | 0.428901 | 3.312 | 6.19E-02 | 14.326 | 1.064 |
| C130 | 102.9715 | 0.527583 | 4.969 | 5.60E-02 | 14.632 | 1.144 |
| C134 | 102.9577 | 0.401421 | 7.453 | 4.93E-02 | 15.522 | 1.444 |
| C135 | 102.9231 | 0.402758 | 7.155 | 5.22E-02 | 15.439 | 1.302 |
| C140 | 102.9886 | 0.429582 | 1.233 | 1.15E-01 | 15.235 | 1.456 |
| C141 | 102.9864 | 0.429705 | 0.4732 | 1.13E-01 | 15.502 | 1.475 |
| C145 | 102.923 | 0.436016 | 3.888 | 1.18E-01 | 15.269 | 1.542 |
| C155 | 102.9931 | 0.469604 | 6.389 | 3.23E-02 | 15.145 | 1.283 |
| C156 | 102.9186 | 0.486715 | 1.656 | 7.47E-02 | 15.210 | 1.251 |
| C179 | 102.9347 | 0.433316 | 0.7846 | 1.71E-01 | 16.105 | 1.618 |
| C181 | 102.9363 | 0.435369 | 7.155 | 4.11E-02 | 15.861 | 1.678 |
| C183 | 102.9523 | 0.436998 | 9.938 | 3.61E-02 | 16.196 | 1.772 |
| C194 | 102.9425 | 0.450717 | 6.389 | 1.00E-01 | 16.347 | 1.718 |
| C215 | 102.9663 | 0.522762 | 7.777 | 3.17E-02 | 15.518 | 1.416 |
| C229 | 102.9449 | 0.415625 | 3.888 | 6.42E-02 | 16.614 | 1.569 |
| C261 | 102.9681 | 0.474545 | 0.8518 | 5.50E-02 | 16.581 | 1.784 |
| C281 | 102.9698 | 0.512495 | 2.129 | 1.07E-01 | 16.845 | 2.205 |
| C295 | 102.9707 | 0.404554 | 1.176 | 1.04E-01 | 17.142 | 2.063 |
| C381 | 102.9212 | 0.511303 | 3.507 | 7.49E-02 | 17.275 | 2.01 |
| C434 | 103.0121 | 0.424351 | 3.650 | 1.01E-01 | 17.992 | 2.433 |
| C487 | 102.9995 | 0.490513 | 3.650 | 9.76E-02 | 17.635 | 1.869 |
| C490 | 103.0467 | 0.4908 | 7.155 | 1.57E-01 | 18.220 | 2.546 |
| C516 | 102.9192 | 0.540043 | 6.168 | 8.26E-02 | 18.010 | 2.357 |
| C578 | 102.9683 | 0.424122 | 0.1827 | 1.88E-01 | 18.894 | 2.234 |
| E46 | 103.0938 | 0.460275 | 2.794 | 4.11E-02 | 13.912 | 0.96 |
| E59 | 103.0862 | 0.444921 | 5.260 | 3.62E-02 | 14.647 | 1.154 |
| E76 | 103.1142 | 0.3916 | 1.555 | 8.20E-02 | 15.192 | 1.292 |
| E97 | 103.1554 | 0.476881 | 6.879 | 4.24E-02 | 15.646 | 1.532 |
| E108 | 103.0696 | 0.533034 | 7.776 | 3.22E-02 | 15.356 | 1.133 |
| E121 | 103.0629 | 0.412029 | 2.235 | 8.96E-02 | 16.199 | 1.562 |
| E134 | 103.1515 | 0.435673 | 0.6125 | 7.25E-02 | 16.243 | 1.652 |
| E139 | 103.0657 | 0.446534 | 8.130 | 3.81E-02 | 16.027 | 1.677 |
| E141 | 103.1073 | 0.450104 | 3.374 | 7.70E-02 | 15.983 | 1.483 |
| E151 | 103.1353 | 0.482273 | 3.888 | 9.18E-02 | 16.603 | 1.923 |
| E167 | 103.1217 | 0.535404 | 6.624 | 4.43E-02 | 16.622 | 1.601 |
| E218 | 103.1614 | 0.474128 | 2.129 | 6.24E-02 | 17.476 | 1.91 |
| E227 | 103.0594 | 0.486915 | 3.888 | 2.03E-01 | 16.998 | 1.708 |
| E231 | 103.1782 | 0.492447 | 9.936 | 1.45E-01 | 17.035 | 2.249 |
| E240 | 103.1178 | 0.512682 | 11.17 | 1.01E-01 | 16.833 | 2.222 |
| E246 | 103.1026 | 0.519228 | 4.968 | 9.29E-02 | 17.315 | 1.747 |
| E327 | 103.0949 | 0.449608 | 2.032 | 5.05E-02 | 17.553 | 2.32 |
| E333 | 103.1613 | 0.454536 | 0.4130 | 2.26E-01 | 18.268 | 2.342 |
| E372 | 103.1799 | 0.50142 | 5.769 | 5.25E-02 | 17.718 | 2.436 |
| E383 | 103.1204 | 0.510435 | 0.4019 | 1.54E-01 | 18.200 | 2.601 |
| N65 | 103.1793 | 0.648279 | 0.1615 | 3.89E-01 | 14.567 | 1.054 |
| N99 | 103.0474 | 0.632065 | 0.5412 | 1.89E-01 | 16.004 | 1.574 |
| N110 | 103.1364 | 0.668296 | 6.140 | 4.78E-02 | 15.173 | 1.373 |
| N116 | 103.085 | 0.685202 | 7.419 | 3.12E-02 | 15.468 | 1.428 |
| N123 | 103.0541 | 0.549887 | 5.237 | 5.25E-02 | 16.239 | 1.153 |
| N127 | 103.0775 | 0.558937 | 8.479 | 1.22E-01 | 16.598 | 2.01 |
| N147 | 103.1077 | 0.587292 | 3.870 | 4.17E-02 | 16.207 | 1.544 |
| N151 | 103.1807 | 0.60032 | 0.7038 | 1.52E-01 | 16.443 | 1.809 |
| N175 | 103.0848 | 0.652286 | 6.594 | 4.25E-02 | 16.721 | 2.023 |
| N203 | 103.0351 | 0.552066 | 10.47 | 4.44E-02 | 17.282 | 1.578 |
| N294 | 103.0349 | 0.554188 | 0.3411 | 1.11E-01 | 17.915 | 2.318 |

TABLE 2 (Continued)

| ID | R.A. | Decl. | Period (days) | Amplitude (mag) | R (mag) | B − R (mag) |
|---|---|---|---|---|---|---|
| N402 | 103.0535 | 0.650805 | 2.225 | 7.66E-02 | 17.605 | 1.995 |
| N413 | 103.0773 | 0.659593 | 8.903 | 9.87E-02 | 17.703 | 2.249 |
| N414 | 103.0995 | 0.660654 | 3.870 | 1.12E-01 | 18.146 | 2.527 |
| N434 | 103.1423 | 0.672277 | 0.3404 | 1.04E-01 | 17.606 | 2.364 |
| N495 | 103.0688 | 0.681308 | 0.6982 | 8.26E-02 | 18.322 | 2.289 |
| N503 | 103.1075 | 0.602628 | 6.359 | 5.65E-02 | 15.843 | 1.791 |
| N616 | 103.0461 | 0.611706 | 2.967 | 1.24E-01 | 18.934 | 2.24 |
| SE39 | 103.0729 | 0.291023 | 4.159 | 4.42E-02 | 14.364 | 0.971 |
| SE69 | 103.2164 | 0.343684 | 7.453 | 5.19E-02 | 15.236 | 1.347 |
| SE142 | 103.1066 | 0.276468 | 2.129 | 2.71E-02 | 16.039 | 1.381 |
| SE159 | 103.1085 | 0.324293 | 1.902 | 5.32E-02 | 16.476 | 1.522 |
| SE173 | 103.1662 | 0.350577 | 5.962 | 2.87E-02 | 16.313 | 1.823 |
| SE206 | 103.0825 | 0.257978 | 2.932 | 7.70E-02 | 16.671 | 1.541 |
| SE222 | 103.0781 | 0.296752 | 4.968 | 7.99E-02 | 16.793 | 1.487 |
| SE225 | 103.1319 | 0.300235 | 9.414 | 4.53E-02 | 16.847 | 2.056 |
| SE234 | 103.1397 | 0.320703 | 5.962 | 9.88E-02 | 16.992 | 2.182 |
| SE239 | 103.1764 | 0.326994 | 4.586 | 7.52E-02 | 17.292 | 1.694 |
| SE262 | 103.1974 | 0.350725 | 6.168 | 5.32E-02 | 17.293 | 1.892 |
| SE284 | 103.1802 | 0.267693 | 0.7644 | 7.32E-02 | 17.410 | 2.18 |
| SE345 | 103.1691 | 0.262441 | 1.305 | 1.11E-01 | 17.929 | 1.782 |
| SE435 | 103.099 | 0.374552 | 2.293 | 9.20E-02 | 17.830 | 2.444 |
| S27 | 102.9716 | 0.34898 | 2.353 | 7.29E-02 | 14.106 | 0.922 |
| S47 | 103.073 | 0.291035 | 4.160 | 6.96E-02 | 14.368 | 0.988 |
| S48 | 102.9683 | 0.296067 | 6.168 | 3.99E-02 | 15.135 | 1.279 |
| S58 | 102.9893 | 0.355676 | 1.296 | 4.51E-02 | 14.788 | 1.187 |
| S64 | 102.9416 | 0.378865 | 5.420 | 3.28E-02 | 14.733 | 1.114 |
| S77 | 103.0076 | 0.288311 | 0.9316 | 1.45E-01 | 15.600 | 1.564 |
| S100 | 102.9752 | 0.371027 | 5.590 | 3.52E-02 | 15.608 | 1.916 |
| S108 | 102.9578 | 0.401381 | 7.453 | 6.24E-02 | 15.547 | 1.443 |
| S128 | 102.9283 | 0.29708 | 4.834 | 8.90E-02 | 16.562 | 1.471 |
| S146 | 102.9338 | 0.334034 | 4.968 | 3.47E-02 | 15.830 | 1.328 |
| S150 | 102.9259 | 0.340571 | 0.7070 | 6.75E-02 | 16.597 | 1.369 |
| S153 | 103.037 | 0.33901 | 3.440 | 3.54E-02 | 15.996 | 1.328 |
| S210 | 103.0113 | 0.314466 | 2.417 | 9.96E-02 | 17.175 | 2.22 |
| S218 | 103.0217 | 0.327381 | 7.155 | 1.25E-01 | 17.069 | 1.999 |
| S257 | 102.9709 | 0.382699 | 10.52 | 7.08E-01 | 16.610 | 1.904 |
| S262 | 102.9817 | 0.388377 | 0.4574 | 1.40E-01 | 16.816 | 1.859 |
| S270 | 102.9707 | 0.404421 | 1.176 | 1.04E-01 | 17.173 | 2.085 |
| S272 | 102.9865 | 0.407135 | 1.671 | 8.52E-02 | 16.486 | 1.911 |
| S274 | 103.0164 | 0.41029 | 3.888 | 1.12E-01 | 17.144 | 2.108 |
| S383 | 103.0163 | 0.37623 | 9.414 | 1.23E-01 | 17.186 | 2.22 |
| S438 | 102.9562 | 0.272712 | 0.3356 | 1.53E-01 | 18.312 | 2.421 |
| S471 | 103.0442 | 0.302407 | 4.968 | 7.49E-02 | 17.867 | 2.403 |
| S553 | 102.9897 | 0.385678 | 8.131 | 1.51E-01 | 18.279 | 2.571 |
| S556 | 102.9506 | 0.388324 | 5.770 | 1.11E-01 | 18.312 | 2.485 |
| W29 | 102.9045 | 0.506024 | 3.501 | 4.36E-02 | 13.5 | 1.012 |
| W54 | 102.7959 | 0.533535 | 3.369 | 3.04E-02 | 14.052 | 1.028 |
| W58 | 102.8065 | 0.55732 | 3.968 | 2.66E-02 | 14.383 | 1.078 |
| W105 | 102.7539 | 0.456523 | 0.4368 | 2.43E-01 | 15.735 | 1.434 |
| W126 | 102.8547 | 0.509649 | 6.613 | 4.89E-02 | 15.528 | 1.523 |
| W145 | 102.8814 | 0.548672 | 7.763 | 3.61E-02 | 15.810 | 1.586 |
| W163 | 102.8515 | 0.574331 | 2.204 | 9.79E-02 | 15.532 | 1.485 |
| W182 | 102.8757 | 0.477998 | 4.960 | 1.09E-01 | 16.528 | 1.866 |
| W198 | 102.8709 | 0.511322 | 5.411 | 3.80E-02 | 16.352 | 1.421 |
| W203 | 102.808 | 0.523545 | 7.142 | 6.59E-02 | 16.858 | 1.977 |
| W205 | 102.8833 | 0.523368 | 1.069 | 8.37E-02 | 16.448 | 1.523 |
| W211 | 102.8133 | 0.541769 | 10.50 | 3.19E-02 | 16.784 | 2.022 |
| W230 | 102.8535 | 0.57583 | 8.503 | 4.76E-02 | 16.721 | 1.973 |
| W269 | 102.8327 | 0.476967 | 11.16 | 4.28E-02 | 17.257 | 2.155 |

TABLE 2 (Continued)

| ID | R.A. | Decl. | Period (days) | Amplitude (mag) | R (mag) | B − R (mag) |
|---|---|---|---|---|---|---|
| W276 | 102.8159 | 0.483833 | 5.580 | 7.01E−02 | 17.579 | 2.318 |
| W282 | 102.7529 | 0.497277 | 5.581 | 9.34E−02 | 17.741 | 2.146 |
| W300 | 102.8834 | 0.512781 | 2.204 | 1.02E−01 | 17.400 | 2.178 |
| W318 | 102.8035 | 0.532219 | 10.50 | 5.52E−02 | 17.622 | 2.366 |
| W330 | 102.8193 | 0.544564 | 4.152 | 1.07E−01 | 17.666 | 1.959 |
| W341 | 102.8919 | 0.559923 | 3.571 | 1.40E−01 | 17.460 | 2.03 |
| W354 | 102.8629 | 0.574649 | 10.50 | 4.73E−02 | 17.322 | 2.283 |
| W356 | 102.9011 | 0.575931 | 3.501 | 7.80E−02 | 17.550 | 1.995 |
| W376 | 102.7992 | 0.60414 | 7.142 | 1.54E−01 | 16.992 | 2.158 |
| W430 | 102.8742 | 0.508038 | 8.928 | 7.15E−02 | 18.016 | 2.527 |
| W440 | 102.8838 | 0.519713 | 8.503 | 7.54E−02 | 18.214 | 2.456 |
| W441 | 102.8596 | 0.520438 | 0.3600 | 1.83E−01 | 18.467 | 2.666 |
| W482 | 102.8181 | 0.578299 | 2.380 | 1.31E−01 | 18.244 | 2.449 |
| W528 | 102.8755 | 0.55438 | 1.174 | 8.61E−02 | 18.332 | 2.558 |
| W535 | 102.8697 | 0.498768 | 4.826 | 3.84E−02 | 14.910 | 1.277 |
| W580 | 102.7727 | 0.482312 | 0.3592 | 1.83E−01 | 18.920 | 2.252 |

W105, C179) are shown in Figure 2. Initially we culled out 140 candidate periodic stars, but two of them are rejected by the eye test. These stars were C23 and C84; their phase plots show an obvious scatter diagram; however, their unphased light curves clearly show that they are variable, likely with much longer periods (if periodic) than the observing run. The majority of the periodic variables have $(B − R) > 1$, with periods from a fraction of a day up to ∼12 days and mostly show amplitude variations of a few hundredths of a magnitude with a few stars up to ∼0.7 mag. Thus we suggest that most of these variables exhibit periodicity caused by axial rotation of a star with a variable degree of nonuniformity of the surface brightness; that is, star spots. Studies of other relatively young open clusters (such as M34, Irwin et al. 2006; M37, Messina et al. 2008, and Hartman et al. 2009) also find that the majority of periodic variables are heavily spotted rotating stars. Thus, since the open cluster NGC 2301 is relatively young (∼100–250 Myr), a large population of rapidly rotating solarlike stars is expected.

A few of the rotational periodic variables are possibly eclipsing binaries, but examination of their light curves was unable to clearly detect primary and secondary eclipses, a shape consistent with an eclipse (given their period), or contained a well enough sampled light curve to be certain of their true nature. We see that the Table 2 entries show a much larger range in $R$ magnitude and orbital period (many multiple day periods) than the stars listed in Table 1.

The columns in Tables 1 and 2 (with the exception of the last column in Table 1) are derived directly from the original observations as presented in H05 and the light-curve examination and analysis performed herein. The R.A. and declination values are J2000. The radial velocity value listed in Table 1 is a crude estimate of the maximum velocity amplitude (the $K$ amplitude) expected for each binary given the assumptions we now list. As a starting point, we have assumed that both stars are of the same mass and the binary has a circular orbit. The mass of the stars were estimated based on their observed $B − R$ value. The last column in Table 1 was produced as an aid to assessing the requirements needed to obtain spectroscopic orbits in future observations.

Figure 3 plots the distribution of periods of our 138 periodic variables. We note that the distribution falls to roughly half its value at half the length of the data set. The periodic variable stars apparent colors range from those of late F stars to late K (see Fig. 4). We note that the selection of periodic variables by color can be useful to separate pulsators and rotation modulations from composite color binaries. The information in Tables 1 and 2 can be used to cull samples of interest from the cluster variables.

Approximately 0.2% of all stars are EBs, giving an expected number of about 5 million systems in the Milky Way Galaxy, of which about 4000 have been discovered and/or catalogued (Guinan 2004). Using this percentage and the assumption that it is correct for an open cluster such as NGC 2301, our light curve samples "should" discover ∼8 EBs. While we report only 5 EB candidates, we are confident that a few more are hiding in Table 2.

## 3. MEMBERSHIP

The detailed analysis of gyrochronological isochrone fits for open clusters requires a good understanding of membership in order to clearly define temporal sequences as well as possible. Previous studies of NGC 2301, including those aimed at membership statistics, have not reached magnitudes as faint as our data set. Even our bright end is poorly sampled by the published literature, with only a few common stars. We do not have the time line or archival data to perform a proper motion or parallax study for our stars, so we will present a couple of qualitative

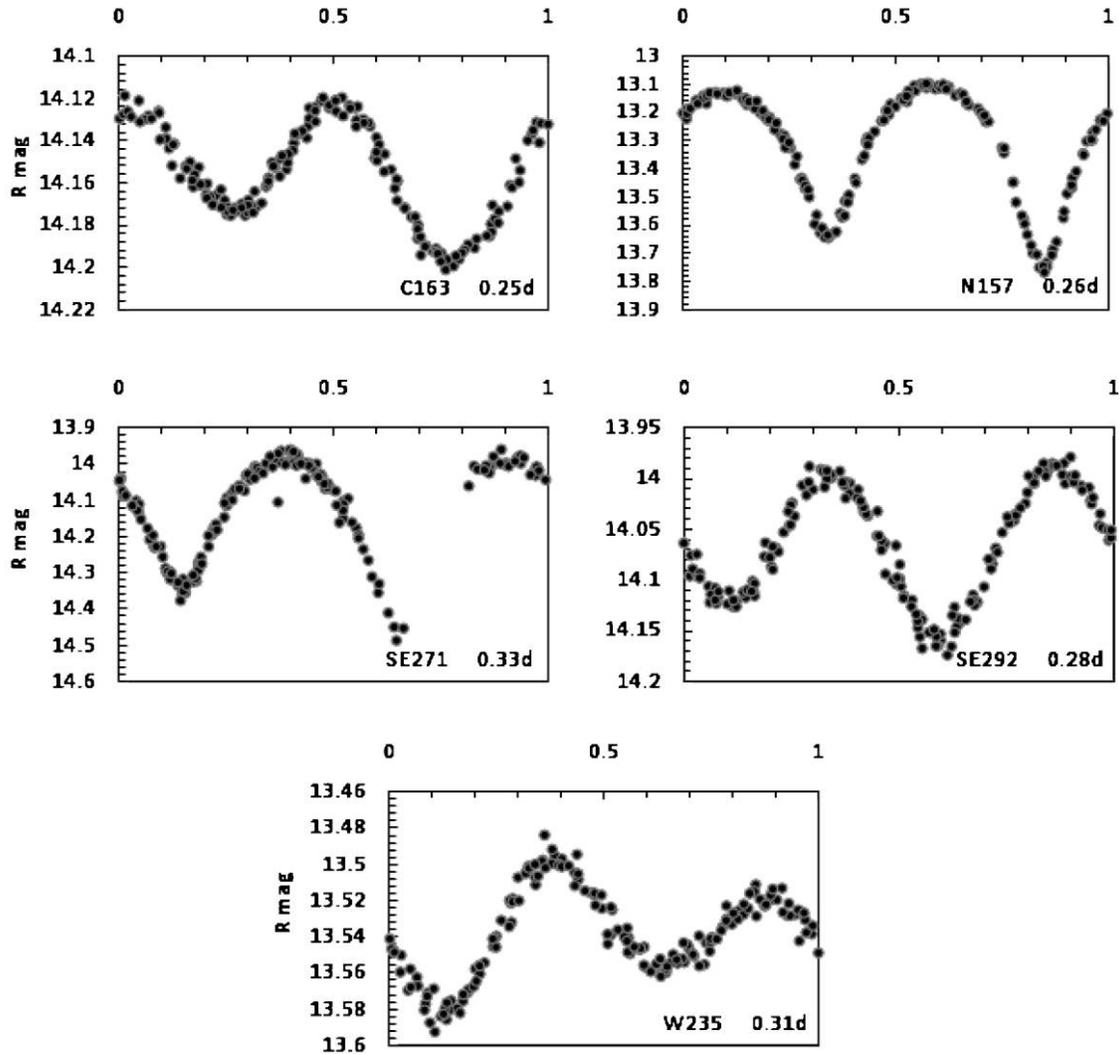

Fig. 1.—Phased $R$-band light curves for five eclipsing binary candidates in NGC 2301. See Table 1. See the electronic edition of the *PASP* for a color version of this figure.

pieces of evidence to support membership for more than half of our 138 periodic variable stars.

First, if we examine the location of the periodic variables on the cluster color-magnitude diagram (see Fig. 4), we note that the majority lie close to or along the main sequence as established in H05. A few lie to the right of and above, or to the left of and below this sequence, being either reddened cluster members or nonmembers, respectively. This "test" is suggestive for cluster membership but not extremely compelling, especially for stars fainter than $R = 17.5$, where photometric color errors are large. Figure 4 also shows 164 Myr and 200 Myr isochrones calculated by Yi et al. (2003), but without additional information on age or membership, the isochrone alone is not a definitive membership criterion. Note too that the CMD isochrones are not good fits for $B - R > 2.1$, where the main sequence deviates up and to the right (see Fig. 4 of H05). We have also marked our 5 EB candidates and our central region stars (refer to Fig. 5) on the diagram. The central region stars are separately analyzed in § 4, as the majority of them lie very close to the isochrones, suggesting that they are very likely to be members. M08 used their kinematic members to set a certain boundary around the isochrone, justifying their photometric members. Spectroscopic study of the NGC 2301 field would be highly desirable to provide a better knowledge of membership by measuring the radial velocity and spectroscopic parallax of the candidate stars.

Second, we present Figure 6a showing a J2000 coordinate map of all 4087 stars from H05 with the 138 periodic variables marked, and Figure 6b showing the distribution of periodic variable stars as a function of radial position. Previous work (e.g., Grubissich and Purgathofer 1962) suggested a cluster radius for NGC 2301 of 15–18′. Since the original field (Fig. 6a) studied by H05 falls within this 18′ region, it is not suggestive for membership. However, the distribution of periodic variables as a function of radial position shows that periodic variable stars have higher degree of concentration than nonperiodic variables or nonvariables around the apparent cluster center (coordinates in § 1). If $n_t$ is the total number of observed stars in the region between circles of radius $r$ and $r + 0.05°$, and $n_v$ is the number of variable stars in that region, we define the radial concentration at that region to be $\rho = \frac{n_v}{n_t - n_v}$. As shown in Figure 6b, the concentration of variables decreases roughly by $1/2$ after $r = 0.1°$.

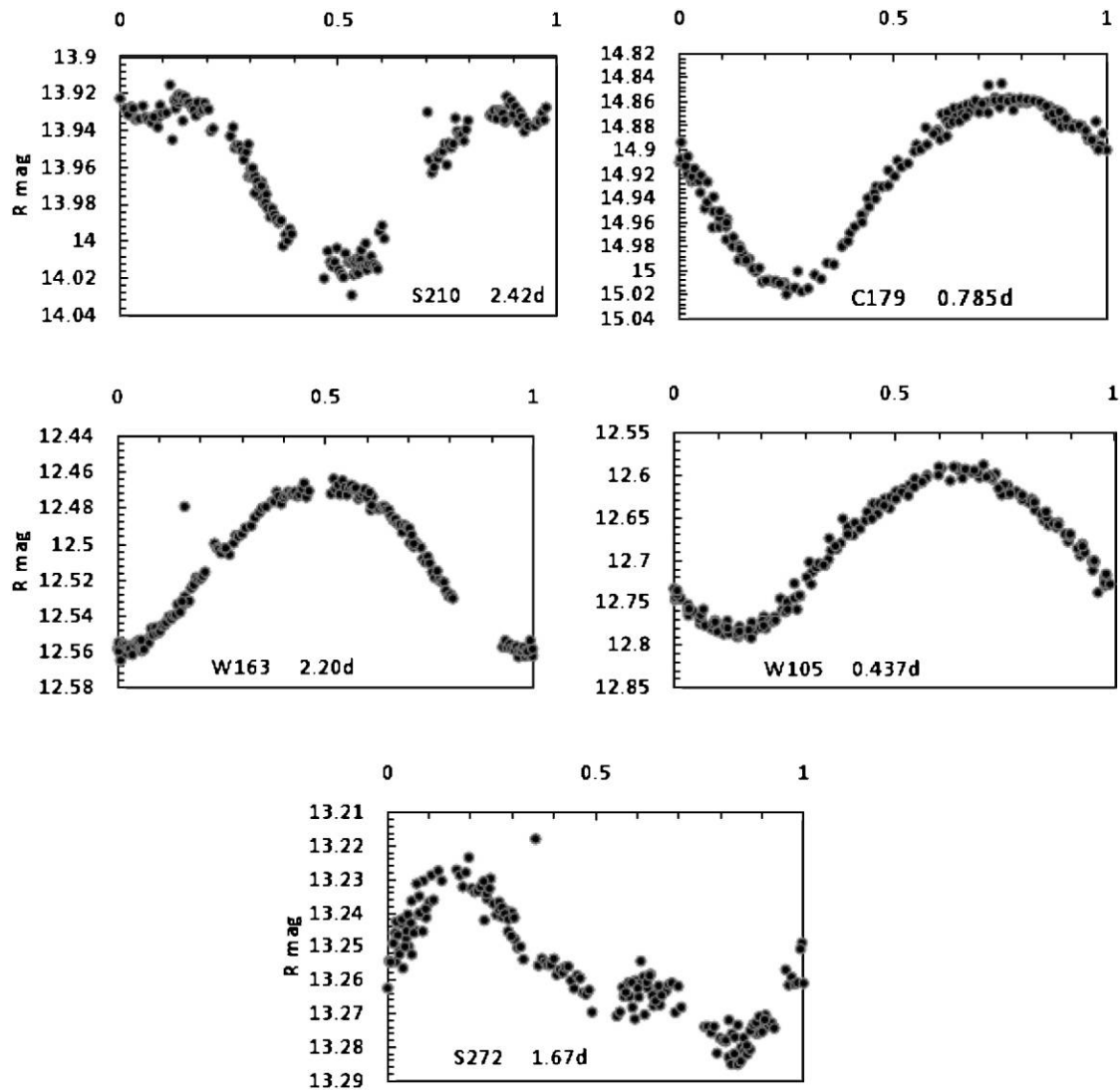

FIG. 2.—Phased $R$-band light curves for four sample periodic variables from our list of 133 additional stars. The phased light curve of S272 is shown as one example of 10 stars that are assumed to be pulsators. See Table 2. See the electronic edition of the *PASP* for a color version of this figure.

Generally the membership becomes more uncertain for fainter stars and those at the outer edges. We see in Figures 4 and 6b that our central region variable stars are quite likely to be true cluster members. This is not a guarantee, however (as noted in H05), as the $r = 11.5$ star C23, located very near the cluster center, was determined to be a background giant.

In order to get a numerical estimate for the background contamination, we have simulated a photometric observation near NGC 2301 using the Trilegal model (Girardi et al. 2005). The contamination for all of our stars is estimated to be ∼43% in the color interval of $0.8 < B - R < 2.3$. Although this value is rather high, field stars are generally old stars, which will not exhibit a rotational variability with short periods and relatively high amplitudes. Therefore the actual contamination to our rotational sample is likely to be much less than 43%.

Taken together, the location of our periodic variables near the main sequence in the cluster color-magnitude diagram, the distribution of periodic variables as a function of radial position and the simulated model of galactic star counts, suggests that from half to $3/4$ of the stars in our periodic sample are highly likely to be cluster members.

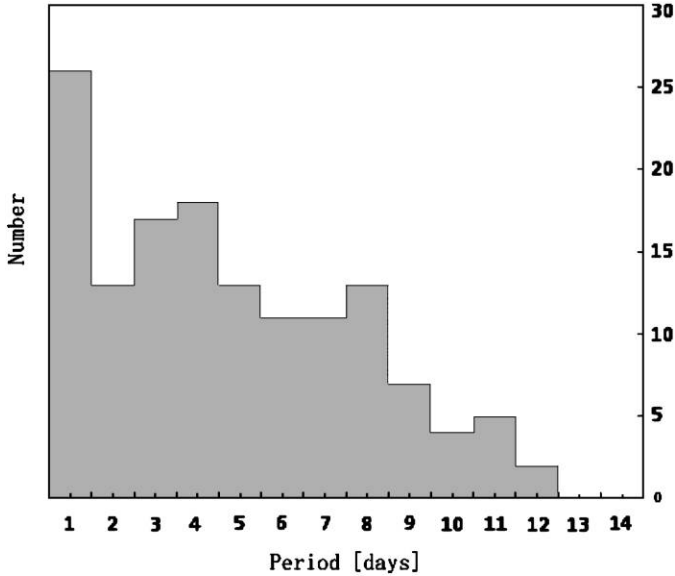

FIG. 3.—Distribution of periods of 138 determined periodic variables illustrating the Nyquist period of $P \sim 8$ days.

## 4. PERIOD-COLOR DIAGRAM ANALYSIS

The analysis of a period-color diagram can be used to estimate the age of an open cluster, with the resulting value being very sensitive to the membership knowledge. Since we have only circumstantial membership understanding, the values that are calculated in this section will likely not be the final word for NGC 2301. However, they are interesting values to compare with previous determinations of the cluster age, which were derived by totally different and independent methods. We also demonstrate the sensitivity of calculated gyrochronological ages based on different assumptions and approaches to the problem.

In this section we analyze our rotation periods and colors in a similar manner to other previously performed works such as M08 and B03. These other studies used $B - V$ color in their analysis, while our data has only $B$ and $R$ measurements. For the comparison, we converted $(B - R)$ to $(B - V)$ using the main sequence colors tabulated in Cox (2000). We calculated a least-square second-order polynomial fit to the color values. Our derived fit, which is shown graphically in Figure 5, is

$$(B - R) = 0.4437(B - V)^2 + 1.2114(B - V) + 0.194. \quad (1)$$

This relation is used for all color conversions in this article.

B03 first analyzed period-color diagrams for several different clusters of different ages and determined two separate sequences of stars. His nomenclature has been used in later studies, thus we will follow that pattern herein. B03 named sequence I (interface) for the diagonal sequence, where the period increased as mass decreased; and the C sequence (convective) for the horizontal sequence of faster rotators. B03 argued that the I sequence stars lose their angular momentum more efficiently than stars on the C sequence, due to their large-scale magnetic fields, where the moment of inertia of the whole star rules the rotational evolution. Stars of the C sequence lose their angular momentum through stellar winds due to their convective magnetic fields, which are not as efficient as the large-scale magnetic fields. Furthermore, B03 claims that a low-mass star will start its main-sequence life on the C sequence, evolving onto the I sequence as it ages and loses its angular momentum due to the magnetic field created by the shear between the envelope and core. In the case of the more massive stars, since they have thin convection envelopes, they leave the C sequence sooner, thus providing the "diagonal" shape to the I sequence

Figure 7a shows the $B - R$ color-period diagram for our 138 identified periodic variables of the cluster. By comparison with the data in M08 and B03, our stars do not fall along two defined bordering sequences representing the two different rotational states; instead they evenly scatter under the "envelope" of the I sequence. An isochrone (discussed in detail later in this section) for a 200 Myr cluster is overplotted on the diagram to reveal the I and C sequences. The envelope starts from $(B - R) \sim 1 (\sim 1.1\ M_\odot)$ and $P \sim 3$ days until $(B - R) \sim 2 (\sim 0.4\ M_\odot)$ and $P \sim 11$ days.

These two sequences are not as well defined in our diagram as for other clusters such as M35 (M08) or Pleiades (B03), where memberships are well established. Assuming the framework suggested by B03 is valid, it is possible that nonmember background field stars contaminate the period-color plane. Although we do not have definite membership knowledge, we can filter out a few stars to help define the sequences as much as possible. One filter is that any star bluer than F8V would not have a convective layer, and thus no small-scale magnetic field to decrease the rotation rate. F8V corresponds to $(B - V) \sim 0.52$ which is $(B - R) \sim 0.94$. We checked our 138 stars, and removed 7 stars with $(B - R) < 1.00$, these were: C96, C85, C79, E46, SE39, S27, and S47. The other assumption was that all of our five EB candidates are real binary systems and were not included in the analysis of the rotation evolution of single stars. Finally, we have checked each phased light curve to see if they have a shape more consistent with a pulsating variable and found 10 such stars (as an example, the phased light curve of S272 is shown in Fig. 2). We assume that all of these 10 stars are true pulsators and did not include them in our gyrochronological analysis using the rotating stars. Thus, we are left with 116 stars to play with.

### 4.1 Gyrochronology

The general forms of the isochrones used to fit data in a period-color diagram are taken from B03. The Skumanich (1972)

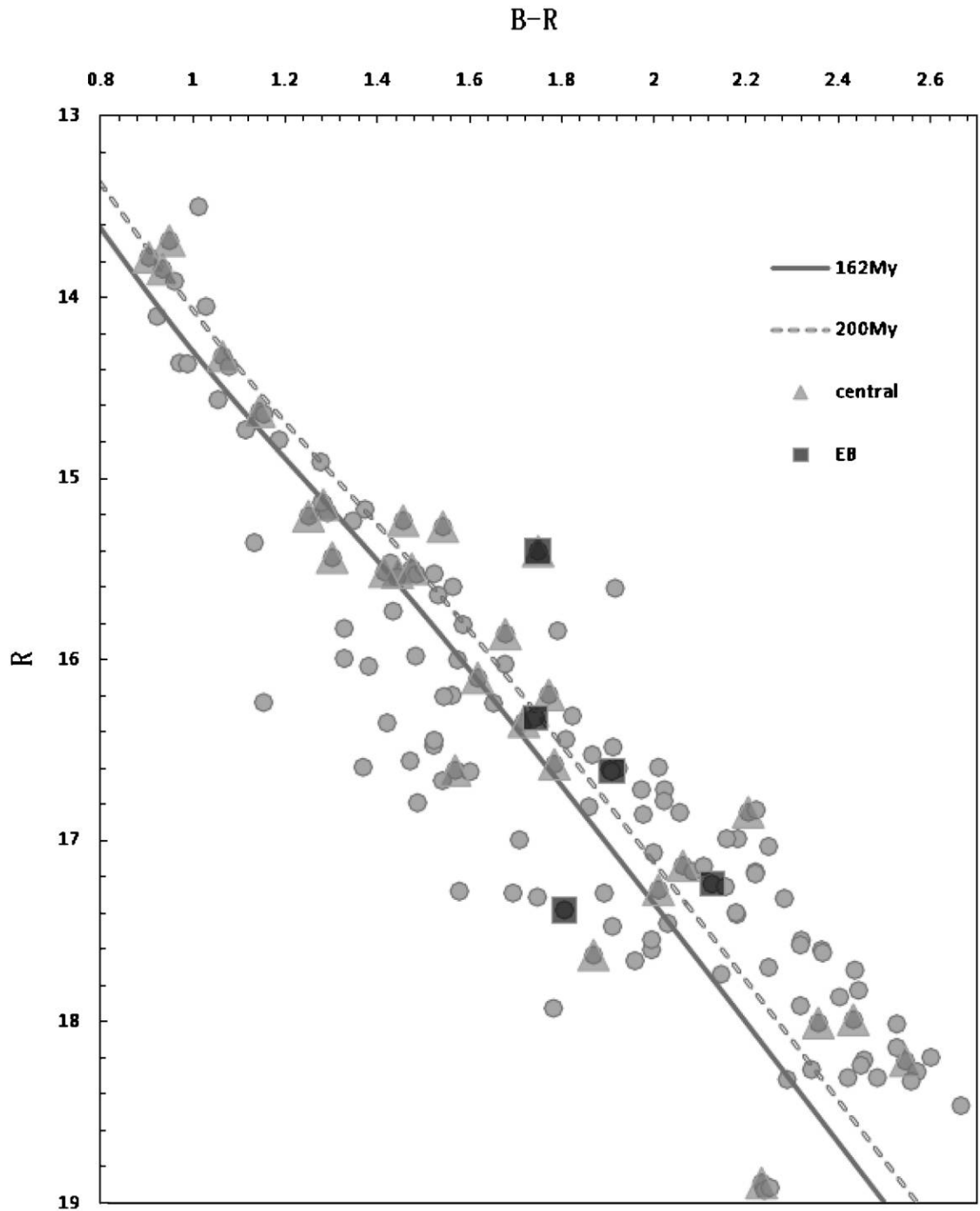

FIG. 4.—CMD of the newly found 138 periodic variables in the open cluster NGC 2301. The two overplotted curves are Yale isochrones with age 162 Myr and 200 Myr (Yi et al. 2003). The five candidate eclipsing binary systems are marked by *squares*, central region stars are marked by *triangles*. See the electronic edition of the *PASP* for a color version of this figure.

relationship of $v \propto t^{-1/2}$ (or $P \propto t^{1/2}$), where $v$ is the rotation velocity of the star, $P$ is the period, and $t$ is the age, has been "tested" by B03 with older Mount Wilson stars and the Sun. B03 concluded that the I sequence stars evolve by this spin-down mechanism. Assuming the Skumanich style spin down governs the I sequence evolution, we adopt the same general relationship as used in B03

$$P = \sqrt{t} \cdot [\sqrt{f(B-R)} - 0.15 f(B-R)], \quad (2)$$

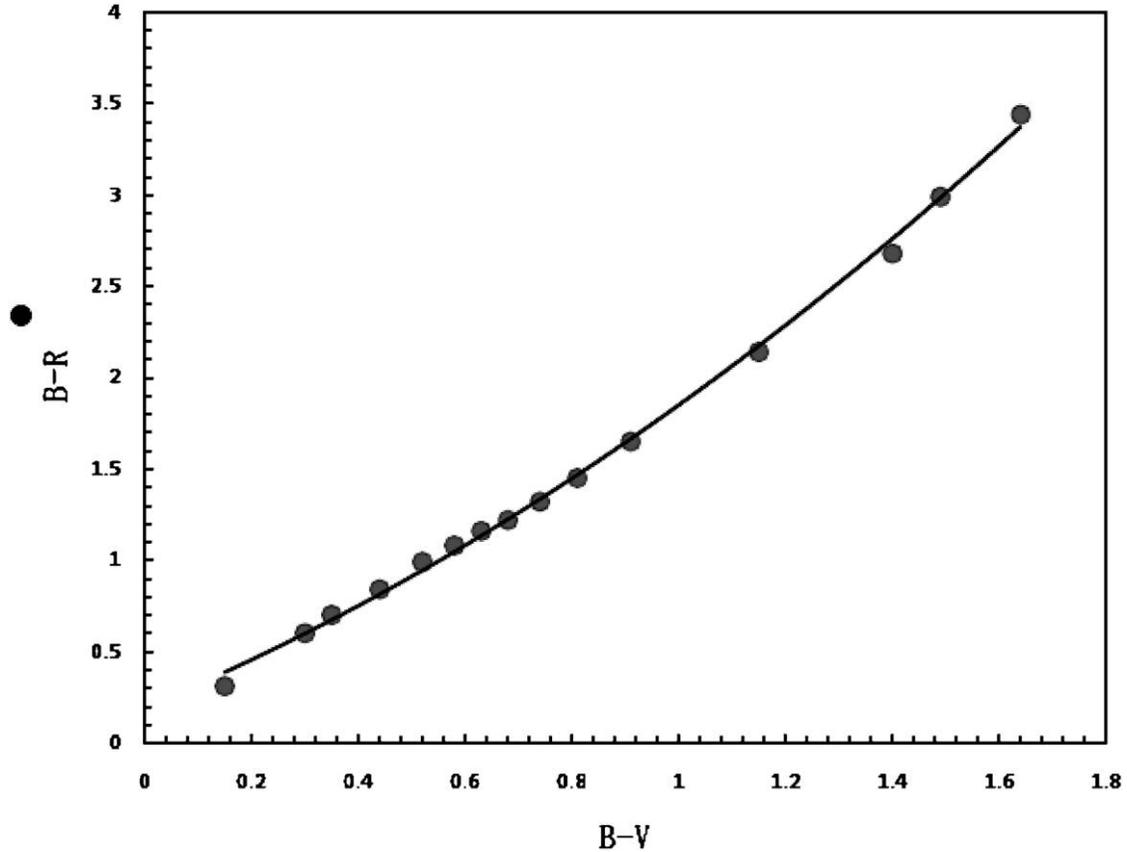

FIG. 5.—$(B-R)$ color as a function of $(B-V)$ color. The data values are from Cox (2000), and the curve is a least-squares parabolic fit, see eq. (1) in text.

where $P$ is the rotation period in days, $t$ is the age in Myr, and we use the $B-R$ color. The function is transformed from the original in B03 by equation (1) in order to represent the color dependency of these data as

$$f(B-R) = -0.055(B-R)^2 + 0.697(B-R) - 0.642.$$

For the isochrones of the C sequence, we also adopt an exponential relationship as in B03, transformed with equation (1) for a different color dependency:

$$P = 0.2 e^{0.01t[-0.055(B-R)^2 + 0.697(B-R) - t/3000]^{-3}}. \quad (3)$$

The relations in equations (2) and (3) are parameterized by the age variable $t$, and thus allow us to build rough isochrones.

There have been previous studies performed to determine the age of NGC 2301. Possibly the earliest determination of an age was 162 Myr (the method is unknown) according to WEBDA. Later Kim et al. (2001) derived the log (age) to be $8.4 \pm 0.1$ Myr (~250 Myr) from the turnoff in the CMD. The most recent number comes from the study of open clusters by Kharchenko et al. (2005), who determined the log (age) to be 8.31, about 205 Myr; however, we note that in their study only a single star was used to make the fit.

Figure 7b shows our 116 stars on the period-color diagram along with isochrones of the 3 previously determined ages. Assuming all 116 stars are members, these 3 fits all look reasonably "good" and at this point we cannot differentiate between any of these age values. The original data set length was ~14 days with the Nyquist period of roughly $P_N \sim 7-8$ days, easily seen in Figure 3 where the distribution falls significantly. Given this short observing length, the cluster may have more rotating members with periods longer than $P_N$. These longer period members (if they exist) would certainly help to better define the I sequence distribution and isochrone fitting.

There are a couple of things we can try to better determine the cluster age. First, we can simply assume all 116 stars are true members, and fit isochrones on subjectively selected stars. This procedure will not give confined age values but will provide interesting values to compare with other ages determined by totally different methods. Second, we can directly apply the Skumanich (1972) relation to some older reasonably "good" established data and analyze it with ours. M08 clearly showed that when they adopted a 625 Myr age for the Hyades cluster data (Radick et al. 1987; Prosser et al. 1995), adopted 150 Myr for M35, and decreased the periods of the Hyades stars by $(625 \text{ Myr}/150 \text{ Myr})^{1/2} \approx 2$ (according to the Skumanich relation), the overplotted Hyades stars provided a good match to the M35 stars, especially in the G dwarf color interval. They also noted that K dwarfs in the Hyades, which are spun up according to the Skumanich relation, all had slower periods than M35 K dwarfs (see Fig. 9, M08). We can use these findings, under certain assumptions, to compare and possibly check the range of previously determined ages for NGC 2301.

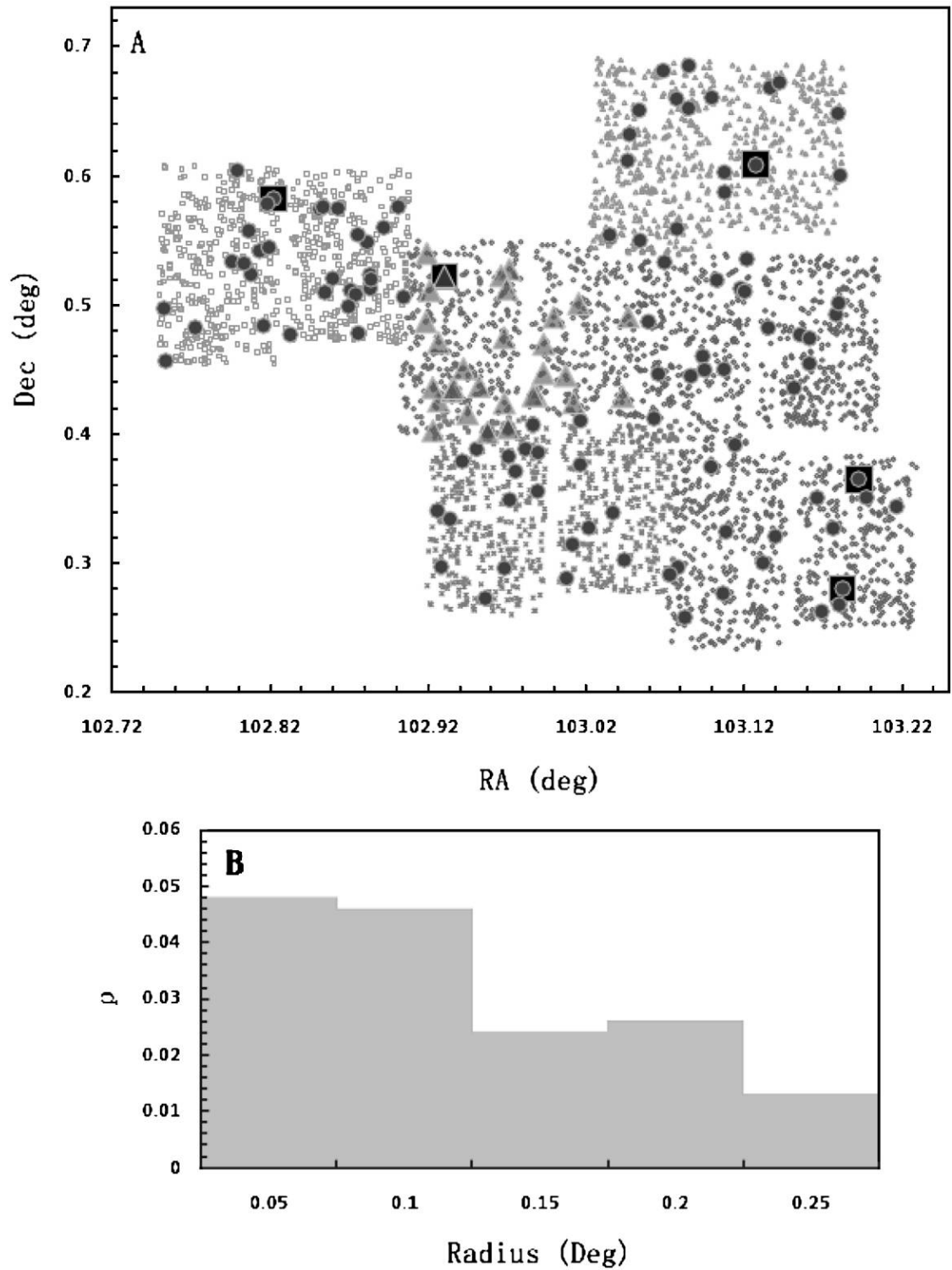

FIG. 6.—(A) The original six OPTIC fields (see Fig. 1 of Tonry et al. 2005) in NGC 2301 with the 138 periodic variables marked. The six fields were designated C, N, E, S, W, and SE; *triangles* mark the spatial location of each of the periodic variables in the central region /C/, and *circles* mark the location for all other variables in other regions. The five EB candidate stars in Table 1 are marked by a larger box. (B) The distribution of periodic variable stars as a function of radial position. The concentration decreases significantly after $r = 0.1°$. See the electronic edition of the *PASP* for a color version of this figure.

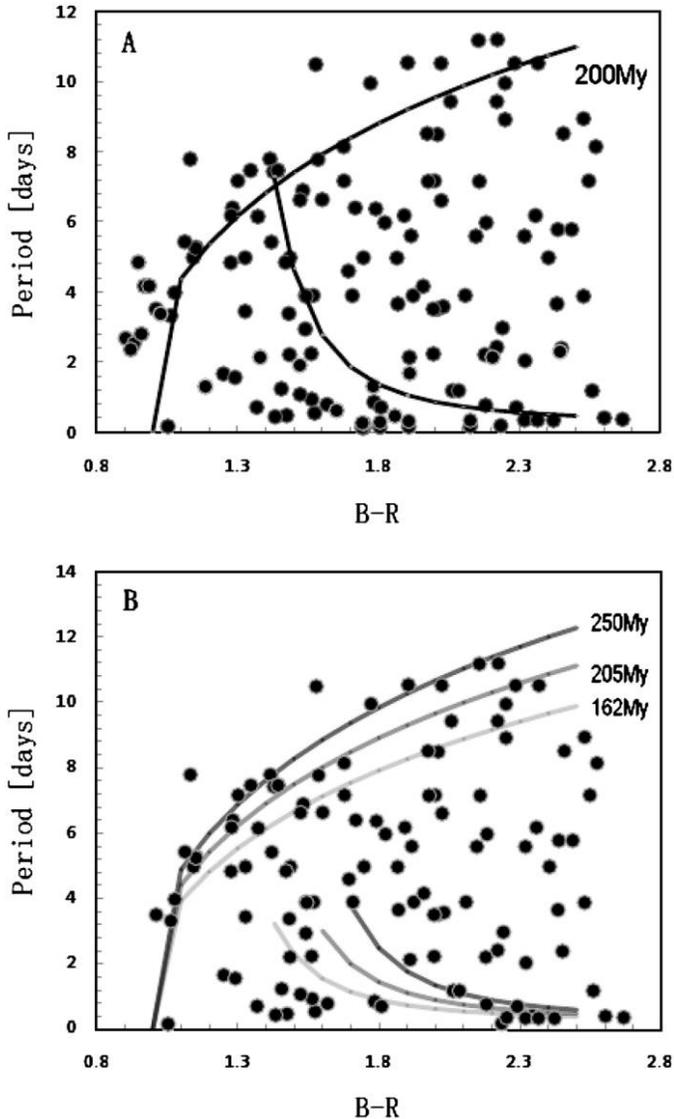

FIG. 7.—(*A*) Period-color diagram of all 138 periodic variables in the field of NGC 2301. An isochrone of 200 Myr is included to reveal I and C sequences.. (*B*) 116 filtered-out stars with 3 gyrochronological isochrones of previously determined ages for the cluster.

Figure 8 shows calculated isochrone fits of the I sequence for our 116 stars. Since the I and C sequences are not clear, with only the I sequence envelope somewhat defined, we made the gyrochronological age fits only for the I sequence. Calculating the least-square fits is totally dependent on what stars we define as I sequence members. Since we have no definite membership information, we have to introduce some kind of subjective criteria. First we defined the I sequence stars as any star located above the criteria line $P = 5.5(B - R) - 4.5$. We used this line for separation of I sequence stars from gap stars, because the $(B - V)$ version of this line was used in M08, a data set with an exceptionally well-defined I sequence. The age of M35 (~150 Myr) is a reasonably close value to the age of NGC 2301 (150–250 Myr), thus the I sequence stars should be close to each other in the period-color plane. This criterion leaves 49 stars to fit the I sequence isochrones. For any data point with coordinates $(B - R)_t$, $P_t$ we define $\Delta$ to be: $\Delta = g[t, (B - R)_t] - P_t$, where $g(B - R)$ is an isochrone function of exactly the same form as equation (3). Then we sum up all $\Delta_t^2$ for each star at different ages $t$, creating a parabolic function $\sum \Delta_t^2 = k(t)$, and the least-square fit occurs at value of $t$ when $dk/dt = 0$.

Figure 8a shows the isochrone fit of $177 \pm 13$ Myr calculated for the 49 stars located above the line segment. Here we note again the assumptions: all stars are members and the criterion line defines the true I sequence stars. To demonstrate the sensitivity of the age, we change the slope of the criterion line to $P = 6.5(B - R) - 4.5$. We choose to increase the slope because according to the most recently determined 2 ages of NGC 2301, it is likely to be older than M35 (~150 Myr). According to the described evolutionary model of these sequences in B03, the I sequence should get steeper as the cluster ages. This new line yields 35 stars to be I sequence members, and the age is calculated to be $212 \pm 15$ Myr. The segment line and the fit are illustrated in Figure 8b. At this point we can narrow down the I sequence stars more by using periods only up to $P_N$. If we adopt $P_N = 8$ days, this removes an additional 10 long-period red stars and leaves 25 stars with periods less than $P_N$ and colors bl'uer than $(B - R) = 1.7$. Figure 8c shows both criteria lines and the calculated age of $196 \pm 14$ Myr. It is of interest that these two final values are very close to the latest determined age value of ~205 My Kharchenko et al. (2005).

The spun-up Hyades stars are overplotted on Figure 9. We assume all of our 116 stars are cluster members and that the Hyades is 625 Myr old. Figures 9a, 9b, 9c show overplots with our 3 different adopted ages for NGC 2301. As demonstrated in M08, if our assumptions are correct, the overplotted spun-up Hyades stars should coincide with the stars of NGC 2301 in the interval of the G dwarfs, which is $0.6 < B - V < 0.8$, and deviate in the interval of K dwarfs, $0.8 < B - V < 1.3$ (see Fig. 9 of M08). If all 116 stars are true members, then the I sequence is determined by the uppermost stars in the G dwarf interval. A simple visual inspection of Figure 9 reveals that ages of 250 Myr and 205 Myr appear to follow the slope of the I

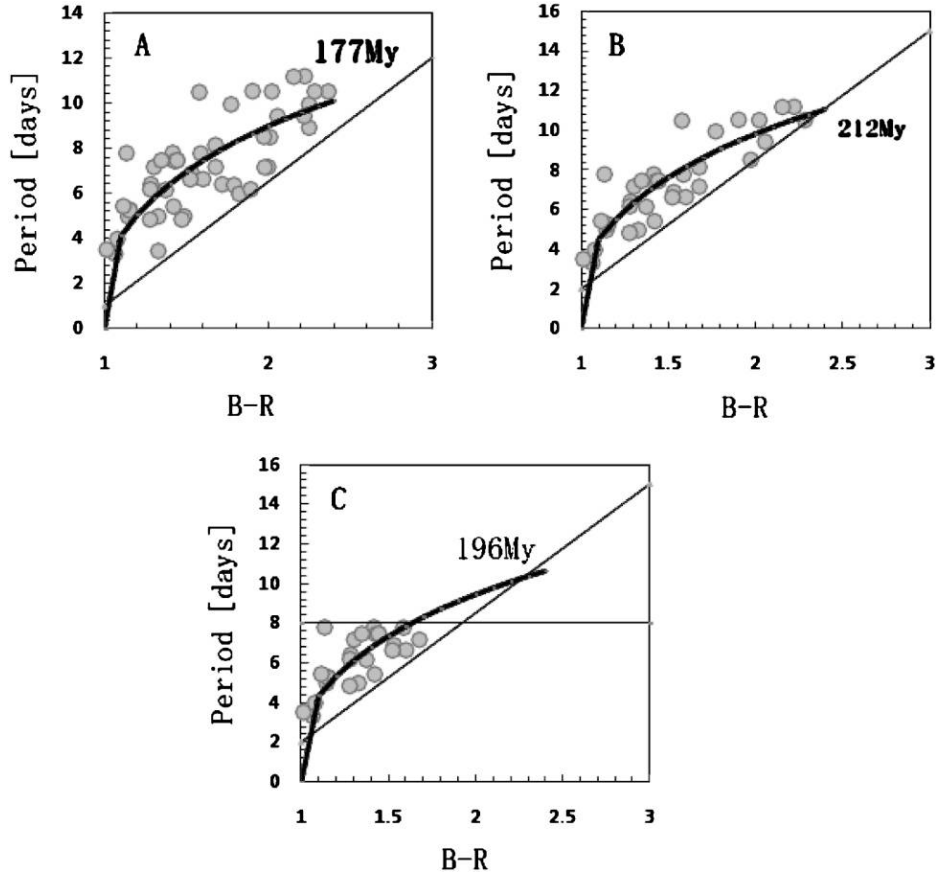

FIG. 8.—(*A*) Least-squares fit isochrone of 177 Myr calculated on 49 stars lying above the line $P = 5.4(B - R) - 4.5$. (*B*) Least-squares fit isochrone of 212 Myr calculated on 35 stars lying above the line $P = 6.4(B - R) - 4.5$. (*C*) Isochrone fit of 196 Myr calculated on 25 stars of the 35 stars in *B*, which have $P < P_N$. All 116 stars are assumed to be cluster members, and all stars separated by these criterion lines are assumed to be the I sequence members in each case correspondingly. See the electronic edition of the *PASP* for a color version of this figure.

sequence over the G dwarf interval more than the 162 Myr age value.

### 4.2 Central Region

An interesting case occurs when we analyze only the stars within the central region (see Fig. 6) of the observed field. Grubissich & Purgathofer (1962) have suggested a cluster radius for NGC 2301 of 15–18′, which is a big circle that contains all of the observed fields by H05 (Fig. 6). However, the period-color plane, CMD, and simulated models of galactic stars suggest that our sample of 116 stars will certainly have some nonmember contamination. The central region of the observed field covers only roughly the cluster center; thus instead of applying a subjective cluster radius value, we culled out all central region stars (as defined in H05) from our list of 116 periodic rotating variables.

Figure 10 shows the period-color diagram for the 23 stars in the central region. The two assumed rotation sequences are still not well defined; however, the I sequence envelope is. We introduce the same criteria lines: stars lying above $P = 6.5(B - R) - 4.5$ are I sequence members and stars with periods shorter than 2 days are assumed to be C sequence members. This time we fit the I sequence gyro-fit along with the C sequence fit.

Part b of Figure 10 shows the same period-color diagram with criteria lines and calculated isochrones curves over plotted. The I sequence isochrone is calculated to be $220 \pm 15$ Myr for 8 stars above the criteria line. This value is in close agreement with previously published values and our gyrochronological analysis discussed above. However, the C sequence isochrone is calculated to be only $96 \pm 10$ Myr on 7 stars with periods shorter than 2 days. At this point it is hard to give any specific reasons without additional membership knowledge. Stellar contamination from background stars and binaries may be the reason. It might even be possible that the C sequence framework is not valid for our case.

### 4.3 Timescales of Migration

The timescales of migration between the two sequences has been investigated initially by B03. Figure 3 of B03 shows the and stars lying above $P = 6.5(B - R) - 4.5$ to be I sequence members; stars in between these two lines are gap region members. Based on our analysis and the previous age determinations, we adopt an age of 220 Myr for NGC 2301. For case 1, there are 45 stars in the gap region, 36 stars in the C sequence region, and 35 stars in the I sequence. For case 2, there are 8 stars counted in the I sequence and gap region respectively, and 7 stars in the C sequence; see Figure 11.

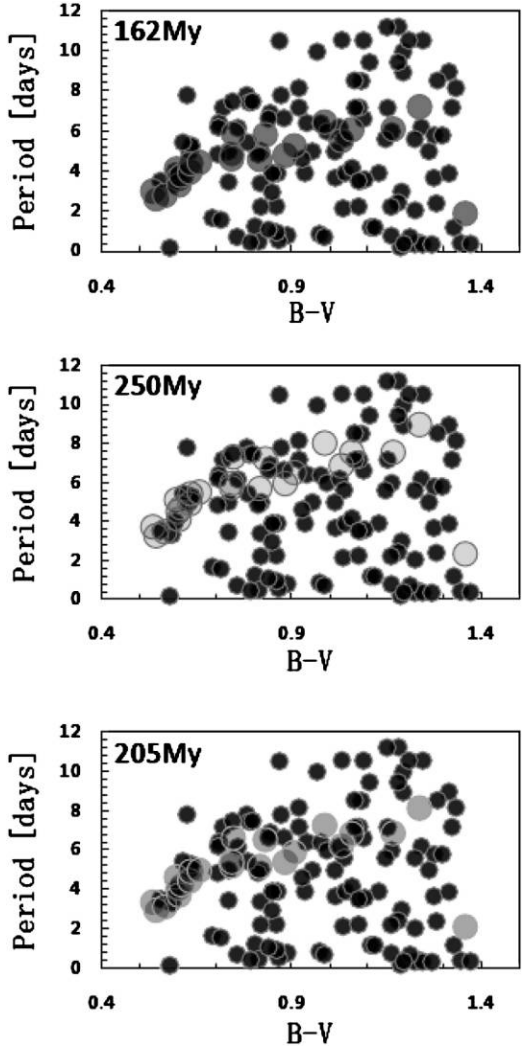

FIG. 9.—Hyades stars spun up according to the Skumanich relation overplotted on period-color diagram of 116 stars in the field of NGC 2301. *Black circles* represent stars of NGC 2301, and larger *gray circles* represent the Hyades stars. All 116 stars are assumed to be cluster members and 625 Myr is adopted as the age of the Hyades cluster. See the electronic edition of the *PASP* for a color version of this figure.

The different approaches of our analysis on the period-color plane result in a best estimate age for NGC 2301 to be in the range of 164–235 Myr, which is roughly the same age range from previous studies (164 Myr–250 Myr). Without definite membership information it is hard narrow it down, however we give more weight to our more limited analysis stars determined by the criteria line and the Nyquist period ($196 \pm 10$ Myr), the analysis of the central region stars ($220 \pm 15$ Myr) and fraction of stars in the different sequences and in the gap region for different age clusters. All stars are in the color interval of $0.5 \leq (B - V) \leq 1.5$. The almost linear trend of this figure suggests that the decrease of population in the C sequence and the increase of population in the I sequence are almost logarithmic relations as a function of time.

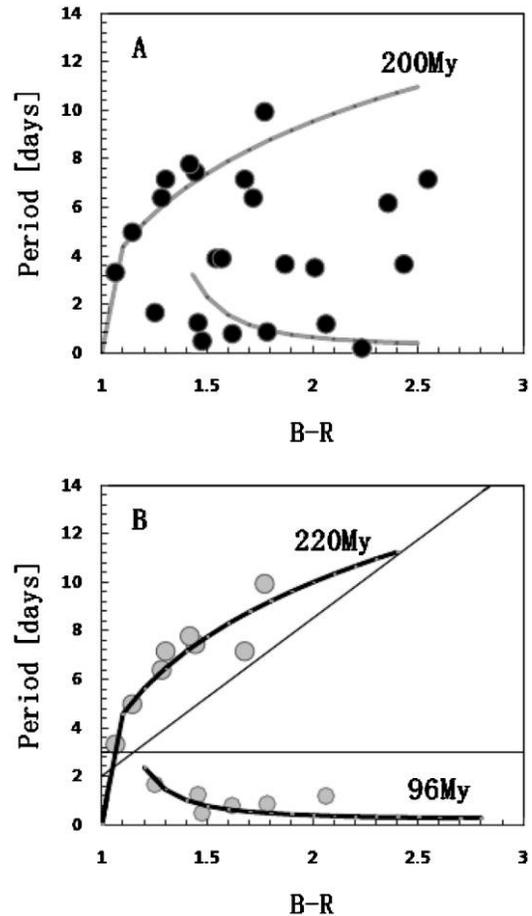

FIG. 10.—(*A*) Period-color diagram of 23 central region stars from the list of 116 candidate rotating variables. An isochrone of 200 Myr age is added. (*B*) The I sequence is separated by $P = 6.5(B - R) - 4.5$ and C sequence by $P = 2$. Calculated gyrochronological ages are 220 Myr for the I sequence and 96 Myr for the C sequence. See the electronic edition of the *PASP* for a color version of this figure.

To see where our data would fall on the diagram in B03, we analyze 2 cases: (1) for all 116 stars, and (2) for only the 23 stars of the central region. For both cases we have used the same criteria as in previous sections to determine the number of stars in the I, C sequences and in the gap region. We assume stars with periods shorter than 2 days to be C sequence members the analysis with spun-up Hyades stars (∼250 Myr). Based on these methods we estimate the age of the cluster to be roughly $210 \pm 25$ Myr.

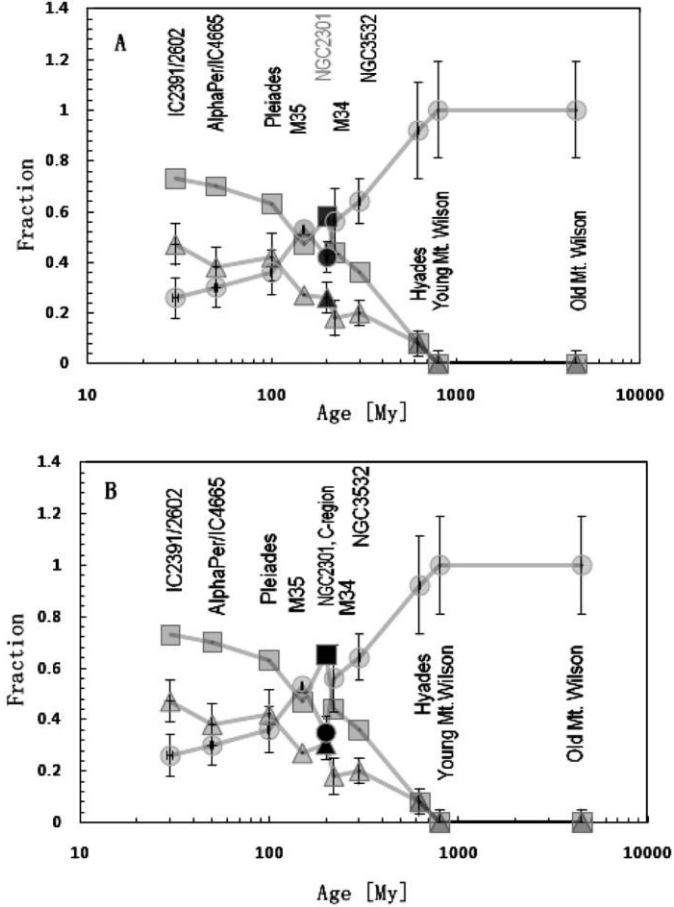

FIG. 11.—Fraction of stars in the I, C sequences and the gap region as a function of time. *Circles* represent fraction of I sequence stars, triangles represent C sequence, and *squares* represent the fraction of gap stars added with C sequence stars. Set of black symbols (square, circle, triangle) show the data for NGC 2301 at the adopted age of 220 Myr. (*A*) Fraction values of all 116 stars in the color interval of $1 < B - R < 2$. (*B*) Fraction values for 23 central region stars in the same color interval as in part *A*. See the electronic edition of the *PASP* for a color version of this figure.

## 5. WHITE DWARF CANDIDATES

In H05, two stars observed in NGC 2301 were suggested as possible white dwarfs. The tentative assignments as white dwarf candidates for the stars N0752 (R.A. = 103.107112, decl. = 0.659648) and C0758 (R.A. = 103.033386, decl. = 0.517640), based on their blue color and their location in the color-magnitude diagram, assumed them to be cluster members (see, e.g., Fig. 5 in H05).

We have photometrically monitored these two stars over the time span from the original observations, 2004 February, through 2006 March with the 0.9 m telescope at Kitt Peak. Table 3 shows their measured $B$ and $R$ magnitudes over this time period, revealing large variability in magnitude (1–1.5 mag) and color changes. These two stars have not again been observed to be as blue as during their initial discovery.

On 2006 January 3 UT, N0752 and C0758 were observed at medium spectral resolution for 10 minutes each (Fig. 12). The observations were made using the Echelette Spectrograph and Imager (ESI; Sheinis et al. 2002) in echelette mode on the Keck II telescope. This setup provided a useful spectral coverage of 4000 to 9500 Å in 10 spectral orders with a constant dispersion of 11.4 km s$^{-1}$ pixel$^{-1}$. A slit width of 1″ was aligned along the parallactic angle providing a plate scale of 0.154″ pixel$^{-1}$ ($R = 3000$–5000 across the optical range). Wavelength calibrations were accomplished using Hg-Ne-Xe and Cu-Ar lamps. The spectrophotometric standard Feige 110 was used for flux calibration. The data were reduced using standard IRAF routines with additional custom analysis routines to properly correct for the large curvature, strong nonlinear photometric response, and small tilt of sky lines. C0758 appears to be a normal looking late G/early K star, possibly of luminosity class III based on its narrow lines. It also shows weak metal lines and may be metal poor. N0752 appears to be a later type star (mid K) with typical looking metal lines but again showing quite narrow lines, possible indicating a luminosity class III object.

It is now clear that these two objects are *not* white dwarfs. Their high level of variability, in particular their color change, are not easily explained by the single spectrum we have obtained. Ideas about binarity with a red and a blue component or some exotic object also seem to be eliminated as the two sources look like, at this epoch, relatively normal single stars.

TABLE 3
Photometric History for N0752 and C0758

| | N0752 | | | C0758 | | |
|---|---|---|---|---|---|---|
| UT Date | $B$ | $R$ | $B - R$ | $B$ | $R$ | $B - R$ |
| 2004 Feb[a] ............. | $17.5 \pm 0.3$ | $18.0 \pm 0.2$ | $-0.55$ | $19.0 \pm 0.3$ | $18.7 \pm 0.2$ | 0.30 |
| 2005 May ............. | $18.0 \pm 0.3$ | $17.5 \pm 0.2$ | 0.50 | $20.3 \pm 0.3$ | $16.5 \pm 0.2$ | 3.80 |
| 2005 Dec ............. | $19.8 \pm 0.3$ | $16.5 \pm 0.2$ | 3.30 | $19.1 \pm 0.3$ | $17.3 \pm 0.2$ | 1.75 |
| 2006 Mar ............. | $17.9 \pm 0.3$ | $16.5 \pm 0.2$ | 1.40 | $19.2 \pm 0.3$ | $17.5 \pm 0.2$ | 1.70 |

[a] Original measurement, Howell et al. (2005)

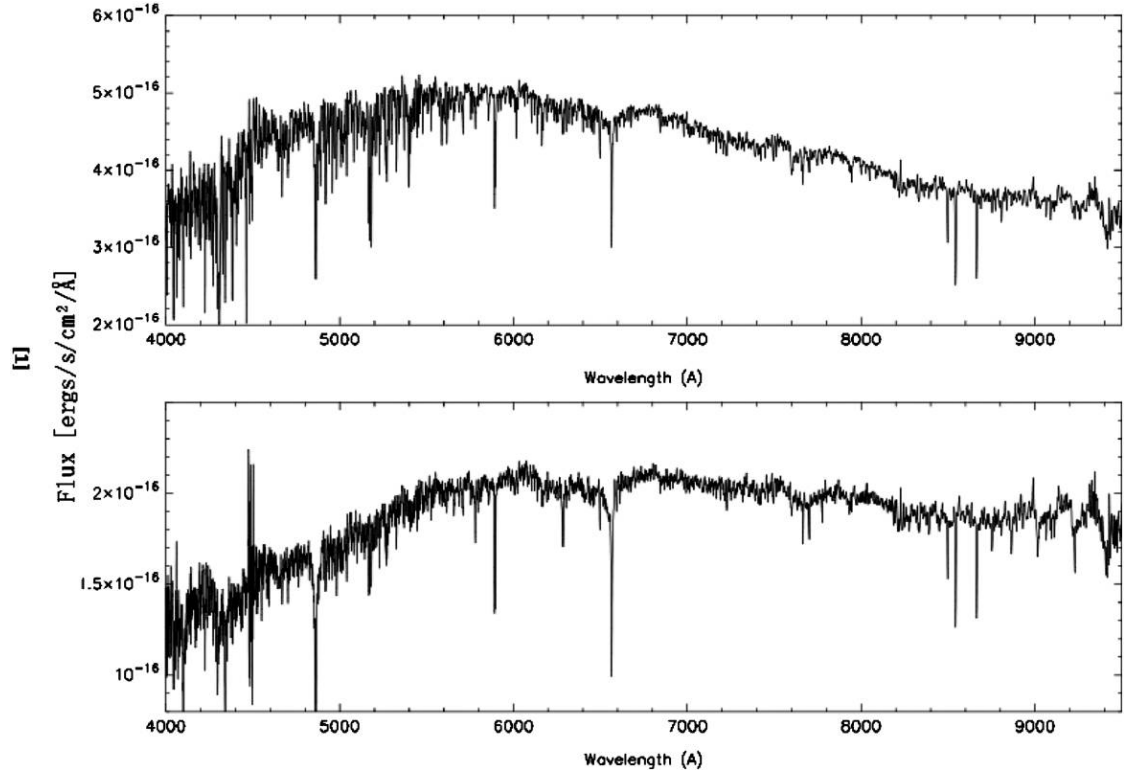

Fig. 12.—Keck spectra of the 2 WD candidates: C0758 and N0752. See text for details.

If they are luminosity class III, they are not cluster members as their apparent magnitudes near $R \sim 17$ will place them far beyond the cluster. Perhaps they are some sort of R CrB type of star semiregularly passing through carbon shell ejection and emission line phases, leading to their variability in brightness and color. If they are cluster members, they would be consistent with being late-type main-sequence stars (most of the time) and one might imagine some sort of flaring activity which may cause their observed blue colors. The true nature of these two point sources will apparently be left as an exercise for the reader.

## 6. CONCLUSION

In the present study we have searched the Howell et al. (2005) data set for the stars with the highest probabilities of periodicity, and discovered 5 eclipsing binary candidates and 133 other new periodic variable stars through a detailed analysis of their light curves. These 133 variable stars are believed to mainly consist of solarlike rotating variables. We attempted to analyze the membership of all the discovered periodic variable stars by their spatial positions on a CMD; by the distribution of periodic variables as a function of radial position, using the stars in the center of the cluster region; and by a simulated photometric observation. We concluded that more than half of our 138 stars are likely to be real cluster members, demonstrating the need for spectroscopic membership determination in the field of this cluster. The WYIN telescope located on Kitt Peak with the HYDRA multifiber spectrograph would be ideal for this work.

On a period-color diagram, we have tested the framework proposed by Barnes (2003). The period-color plane of our data did not show temporal I and C sequences, but it showed the envelope of the I sequence. Assuming the Barnes (2003) framework as valid, we have analyzed the previously determined ages for NGC 2301 (all found through a different method) by the heuristic isochrones. These ages also have been tested using spun-up Hyades cluster stars over plotted on NGC 2301 data. We applied a nonlinear least square fit of the rotational isochrones of our data and found ages in the range of 164–235 Myr depending on different subjective criteria. Based on our total analysis, we estimate the age of NGC 2301 to be roughly $210 \pm 25$ Myr, a value in a good agreement with 205 Myr found by Kharchenko et al (2005). Finally we have nullified two white dwarf candidates in the field of the open cluster NGC 2301 based on spectroscopic observations.

All 4078 photometric light curves of NGC 2301 are available by contacting the authors, or online at WEBDA (see footnote 1), or at NeXSci through the NASA Star and Exoplanet Database (NStED).[2]

---

[2] At http://nexsci.caltech.edu/archives/nsted/.

Some of the data presented herein were obtained at the W. M. Keck Observatory, which is operated as a scientific partnership among the California Institute of Technology, the University of California, and the National Aeronautics and Space Administration. The Observatory was made possible by the generous financial support of the W. M. Keck Foundation. The authors wish to recognize and acknowledge the very significant cultural role and reverence that the summit of Mauna Kea has always had within the indigenous Hawaiian community. We are most fortunate to have the opportunity to conduct observations from this mountain. Also this research has made use of the WEBDA database, operated at the Institute for Astronomy of the University of Vienna.

The authors also wish to thank Mark Huber, Robert H. Becker, and Richard L. White for obtaining the Keck II EIS spectra; Con Deliyannis and Daryl Willmarth for obtaining some of the photometric observations; Dr. Sydney Barnes for providing data and giving valuable advice to this research project; and to an anonymous referee for helping to improve this article.